\documentclass{ws-ijmpa}
\usepackage[super,compress]{cite}
\usepackage{graphicx}
\usepackage{hyperref}

\usepackage{ulem}

\newcommand{\T}[2]{\ensuremath{{\bf #1}}_{#2\text{\scriptsize T}}}
\newcommand{\TT}[2]{\boldsymbol{#1}_{#2\text{T}}}

\newcommand{\Tsc}[2]{#1_{#2\text{T}}}
\newcommand{\Tscsq}[2]{#1^2_{#2\text{T}}}

\newcommand{\no}{\nonumber \\}
\newcommand{\parz}[1]{\ensuremath{\left(#1\right)}}
\newcommand{\order}[1]{\ensuremath{O\parz{#1}}}

\newcommand{\xbj}{\ensuremath{x}}

\newcommand{\diff}[1]{\mathrm{d}#1}
\newcommand{\msbar}{\ensuremath{\overline{\rm MS}}}

\newcommand{\erefs}[2]{Eqs.~(\ref{e.#1})--(\ref{e.#2})}
\newcommand{\eref}[1]{Eq.~(\ref{e.#1})}

\newcommand{\sref}[1]{Sec.~\ref{s.#1}}

\begin{document}
\markboth{Ted Rogers}{Transverse moments of TMD parton densities}

%
\catchline{}{}{}{}{}
%
\title{Transverse moments of TMD parton densities and ultraviolet divergences\footnote{Originally prepared for DIS2020, canceled due to the Covid-19 epidemic.}}

\author{Ted Rogers\footnote{Electronic address: trogers@odu.edu, \href{https://orcid.org/0000-0002-0762-0275}{ORCID: 0000-0002-0762-0275}} 
}

\address{Jefferson Lab, 12000 Jefferson Avenue, Newport News, VA 23606, USA  and,\\
Department of Physics, Old Dominion University, Norfolk, VA 23529, USA \\
trogers@odu.edu}

\maketitle

\begin{history}
\received{28, July, 2020}
\end{history}

\begin{abstract}
I review some open questions 
relating to the large transverse momentum divergences in transverse moments of transverse momentum dependent (TMD) parton correlation functions. I also explain, in an abbreviated and summarized form, recent work that shows that the resulting violations of a commonly used integral relation are not perturbatively suppressed. I argue that this implies a need for more precise definitions for the correlation functions used to describe transverse moments. \\
\\
JLAB-THY-20-3227
\keywords{QCD; Perturbation Theory; Transverse Momentum Dependence.}
\end{abstract}

\section{Introduction}
\label{s.observables}

Efforts to study the partonic structure of nucleons have  motivated a search for physical observables with specific sensitivity to intrinsic nonperturbative parton transverse momentum. 
Classic examples are cross sections differential in a small final state transverse momentum. In the discussions below, I will use semi-inclusive deep inelastic scattering (SIDIS), differential in the transverse 
momentum $\T{P}{h}$ of the measured hadron, 
\begin{equation}
 \frac{\diff{\sigma^{{\rm SIDIS}}}{}}{\diff{x} \diff{y} \diff{z} \diff{^2\T{P}{h}}} \, , \label{e.sidis}
\end{equation}
as a reference process since it is a typical example with relevance to several upcoming experiments. 

When access to the detailed behavior of \eref{sidis} or similar processes at small transverse momentum 
is not practical or convenient, it can be useful to instead work with other, simpler observables that nonetheless retain sensitivity to intrinsic transverse momentum. 
For example,  
integrating with a power $n$ of a transverse momentum component $\alpha$ retains sensitivity to polarization 
effects associated with TMD parton density functions (PDFs) and fragmentation functions (ffs).  These weighted moments,
\begin{equation}
\int \diff{^2\T{P}{h}}{} \parz{\Tsc{P}{h}^\alpha}^n \frac{\diff{\sigma^{{\rm SIDIS}}}{}}{\diff{x} \diff{y} \diff{z} \diff{^2\T{P}{h}}} \, , \label{e.weighted}
\end{equation}
with $n > 0$, are related to spin asymmetries and are 
useful for accessing the large variety of possible correlations between intrinsic transverse momentum and 
spin~\cite{Tangerman:1994bb,Mulders:1995dh,Bacchetta:2006tn} while preserving the simplicity of a  transversely integrated quantity. In a parton model with TMD PDFs, the polarization dependent TMD PDFs 
appear in \eref{weighted} as weighted transverse moments of the TMD PDFs~\cite{Boer:2003cm}, as will be discussed below. 

Aside from simplicity, there are additional theoretical advantages to working with transversely weighted moments of TMD correlation functions instead of transversely differential cross sections. Descriptions of \eref{weighted} can make use of theoretically derived relationships between transverse moments of \emph{TMD} correlation functions 
and (sometimes higher twist) \emph{collinear} correlation functions~\cite{Boer:2003cm}. The most basic integral relation of this kind connects the zeroth transverse moment 
of an unpolarized quark TMD PDF $f_1(x,\Tsc{k}{})$ to the unpolarized collinear quark PDF $f(x)$,
\begin{equation}
\int \diff{^2 \T{k}{} } \; f_1(x,\Tsc{k}{}) = f(x) \, , \label{e.tmd2coll}
\end{equation}
which follows directly from a number density interpretation of PDFs. 
Many other integral relations analogous to \eref{tmd2coll} have been proposed for other types of TMD correlation functions, usually  involving some combination of weighted transverse moments and higher twist collinear correlation functions. 
(This basic idea will be further reviewed below.)
Applied to \eref{weighted} with $n > 0$, these integral relations hint that effects normally associated with TMD PDFs might be accessed through a kind of hybrid of higher-twist-collinear and TMD factorization theorems, and that the advantages of one 
or the other might be exploited according to the needs of a particular effort. Within this view, higher twist collinear correlation functions and TMD correlation functions are seen as different ways of representing the same (or roughly the same) underlying physics.

One specific situation where this idea has been put into practice  
is in the treatment of observables for which standard TMD factorization derivations fail to hold, such as in the 
production of hadrons in hadron-hadron collisions~\cite{Collins:2007nk,Rogers:2010dm}.  Here, integral relations that connect 
TMD and collinear correlation functions are used to relate transverse spin asymmetries in $H_1 + H_2 \to H_3 + X$ collisions to TMD functions like the Sivers function in SIDIS~\cite{Gamberg:2010tj,Kang:2011hk,Kang:2012xf}, thus providing ways to compare intrinsic transverse momentum effects across a variety of experimental settings while side-stepping  
complications with factorization that can arise in some processes when intrinsic transverse momentum is involved. 

Another application of the theory of transverse moments is their use in 
equations of motion and Lorentz invariance relations~\cite{Mulders:1995dh,Lorce:2014hxa,Metz:2008ib,Kanazawa:2015ajw}. These are systems of equations that connect the large number of partonic correlation functions that can contribute to an observable, thereby reducing the number of parameters needed for modeling or in phenomenological extractions. They have been incorporated into many of the models currently used for phenomenological applications, for example in Refs.~\citen{Avakian:2010br,Lorce:2014hxa,Gamberg:2017gle,Bastami:2018xqd,Bastami:2020asv}. Most of these Lorentz invariance and/or equations of motion relations involve a mixture of 
collinear correlation functions and weighted moments of TMD correlation functions. 

However, the mix of collinear and TMD ingredients in descriptions of observables like \eref{weighted} raises 
questions concerning exactly which form of factorization is relevant to a given situation. One problem is that integral relations that include \eref{tmd2coll} and their transverse moment analogs involve ultraviolet divergences that ultimately need to be regulated within some scheme choice.  
This connects naturally to questions about the optimal treatment of QCD evolution in weighted 
observables. For processes that are inclusive in transverse momentum, the  
$Q^2$-dependence is associated with the integration of  transverse momentum up to very large values, of order $Q$. A question, then, is whether an observable like 
\begin{equation}
\label{e.evolving}
\frac{\diff{}}{\diff{\ln Q^2}} \parz{ \int \diff{^2\T{P}{h}}{} \parz{\Tsc{P}{h}^\alpha}^n \frac{\diff{\sigma^{{\rm SIDIS}}}{}}{\diff{x} \diff{y} \diff{z} \diff{^2\T{P}{h}}} } \, ,
\end{equation}
is governed mainly by the standard renormalization group techniques of collinear factorization, as might normally be expected for something inclusive in transverse momentum, or whether the known subtleties of TMD factorization (including rapidity divergences, non-perturbative evolution, etc) need to be taken into account, as might be expected for an observable sensitive to intrinsic transverse momentum. Notice that a modification of the large $\Tsc{P}{h}$ behavior in Eq. (4), which can enter theoretical calculations through regulators or cutoffs on large partonic $k_T$, affects the scale dependence when that modification depends on the kinematics of the process. Of course, if the range of integration is chosen to cover the entire kinematically allowed region, it will depend on kinematical variables like $Q^2$. Thus, the evolution in \eref{evolving} depends on the details of how different regions of transverse momentum are partitioned and included or excluded in the integration.

Below I will argue that these questions are 
more subtle than they might appear at first sight. When extended to transverse moments, relations analogous to \eref{tmd2coll} have potentially large (perturbatively unsupressed) violations from the integration into large $\Tsc{k}{}$. Before discussing this, in \sref{sec1} I will present a more detailed overview of the different types of transverse momentum that enter into integrals like \eref{tmd2coll}. In 
\sref{weighted}, I will extend that discussion to transverse moments of correlation functions. I will review the more interesting subtleties that can arise for weighted correlation functions in \sref{tmdren}. In \sref{summary} I will discuss some proposals for how they might be dealt with in practice.

\section{Sensitivity to intrinsic vs. large transverse momentum}
\label{s.sec1}

For classifying transverse momentum in \eref{sidis}, it is convenient to use $\Tsc{q}{} = | \T{P}{h,} |/z$, so this is the transverse variable I will use from here forward. 

TMD factorization is valid in the region of
transverse momentum much smaller than the hard scale ($\Tsc{q}{} \ll Q$). 
When $\Tsc{q}{}$ is comparable to the hard scale $Q$, the $\Tsc{q}{}$-dependence no longer factorizes
into separate TMD functions of $x$ and $z$.
However, this large $\Tsc{q}{}$-dependence 
is, in principle, perturbatively describable in a purely collinear factorization treatment, with all the transverse momentum generated directly in the hard, perturbative (but generally process-dependent) subprocess. The full description of the cross section across all 
$\Tsc{q}{}$ involves a TMD-based description at small $\Tsc{q}{}$ combined with a collinear-based description at large $\Tsc{q}{}$, and these two separate factorization treatments  
need to be merged in a consistent way to achieve an accurate point-by-point description of the cross section. 
This is typically implemented with an additive modification (a ``$Y$-term'') to the familiar TMD description, which I will write here in abbreviated form in 
terms of the hadronic tensor as
\begin{align}
W(\xbj,z,\T{q}{})^{\mu \nu} =W(\xbj,z,\T{q}{})^{\mu \nu}_\text{TMD} + Y(\xbj,z,\T{q}{})^{\mu \nu}     \, .  
 \label{e.WpY}
\end{align}
The first term, $W(\xbj,z,\T{q}{})^{\mu \nu}_\text{TMD}$, has the familiar structure 
of most TMD treatments (e.g., Ref.~\citen{Mulders:1995dh}) -- there is a hard partonic tensor $\hat{W}(Q^2)^{\mu \nu}$ and a convolution of TMD functions:
\begin{align}
&{} W(\xbj,z,\T{q}{})^{\mu \nu}_\text{TMD} \no &{}
\; \equiv \hat{W}(Q^2)^{\mu \nu} \int \diff{^2 \TT{k}{1}}{} \diff{^2 \TT{k}{2}}{} \delta^{(2)}(\TT{k}{1} + \T{q}{} - \TT{k}{2}) 
f_1(x,\Tsc{k}{1}) D(z,z\Tsc{k}{2}) \, .  
 \label{e.Wterm}
\end{align}
The function $f_1$ is an unpolarized TMD PDF and $D$ is a TMD fragmentation function. 
For brevity, I will continue to suppress flavor and Dirac indexes, polarizations, and auxiliary arguments like 
renormalization scales. The second term in \eref{WpY}, $Y(\xbj,z,\T{q}{})^{\mu \nu}$, is the modification necessary to account for the large transverse momentum region ($\Tsc{q}{} \sim Q$). A precise definition 
for $Y(\xbj,z,\T{q}{})^{\mu \nu}$ can be found in many places,  see, for example, Ref.\citen{Collins:2016hqq} and references therein, although the details are unimportant here. What matters for the present discussion is only  
that $Y(\xbj,z,\T{q}{})^{\mu \nu}$ is perturbatively calculable in collinear factorization, but is not factorizable into separate $x$ and $z$ dependent TMD correlation functions.   
It starts at order-$\alpha_s$ and accounts for the process-specific, nonfactorizable transverse momentum dependence that can
arise at large $\Tsc{q}{}$. 

The tail behavior at large $\Tsc{q}{}$ starts at order $\alpha_s$, so it is reasonable 
to first try approximating integrals over transverse momentum 
by neglecting $Y(\xbj,z,\T{q}{})^{\mu \nu}$ and assuming
\begin{equation}
W(\xbj,z,\T{q}{})^{\mu \nu} \approx W(\xbj,z,\T{q}{})^{\mu \nu}_\text{TMD} \, . \label{e.pm}
\end{equation}  
The zeroth transverse moment of the hadronic tensor then becomes
\begin{align}
&{} \int \diff{^2 \T{q}{}} W(\xbj,z,\T{q}{})^{\mu \nu} \no
&{} \qquad \approx  \hat{W}^{\mu \nu} \parz{\int \diff{^2 \TT{k}{1}}{} \, f_1(x,\Tsc{k}{1})  } \parz{\int \diff{^2 \TT{k}{2}}{} \, D(z,z\Tsc{k}{2})  } \no
&{} \qquad = \hat{W}^{\mu \nu} f(x) D(z) \, , \label{e.reduction}
\end{align}
where the second line has used \eref{tmd2coll} and 
its analog for the fragmentation function $D(z)$. Here, I have had to ignore the ultraviolet divergences discussed in \sref{observables}, but if I permit this I recover the natural expectation for the parton model on the last line of  \eref{reduction}. Ultimately, of course, it is important to go beyond this and to correct for the ultraviolet ambiguity created by the divergent behavior. The appearance of divergent integrals suggests that careful attention needs to be paid to the details of 
the underlying operator definitions for the correlation functions. I will revisit this point after repeating the above discussion for the more interesting case of weighted observables in the next section. 

Of course, the true cross section has a maximum kinematical $\T{q}{}$, and the $\T{q}{}$ integral is finite. This can be seen in the full factorization formula, \eref{WpY}, for the cross section for all $\T{q}{}$. 
Integrated over transverse momentum, it is 
\begin{equation}
\label{e.fullintegral}
\int \diff{^2 \T{q}{}} W(\xbj,z,\T{q}{})^{\mu \nu} =  \int \diff{^2 \T{q}{}} W(\xbj,z,\T{q}{})^{\mu \nu}_\text{TMD} +   \int \diff{^2 \T{q}{}} Y(\xbj,z,\T{q}{})^{\mu \nu} \, .
\end{equation} 
If the integrals extend to infinity, the first term on the right-hand side is divergent 
just as in \eref{reduction}. However, the 
$Y(\xbj,z,\T{q}{})^{\mu \nu}$ term contains an equal and opposite ultraviolet divergence, assuming it is constructed via the usual subtraction procedure, so the details of 
any ultraviolet regulators cancel between the $W(\xbj,z,\T{q}{})^{\mu \nu}$ and $Y(\xbj,z,\T{q}{})^{\mu \nu}$ terms. 

Writing the $\T{q}{}$-integration as in \eref{fullintegral} highlights the fact that large-$k_T$ divergences in integrals like \eref{tmd2coll} are symptoms of having neglected the $\T{q}{} \sim Q$ behavior in the $Y(\xbj,z,\T{q}{})^{\mu \nu}$ term in approximations like \eref{pm}. The ambiguities introduced by  large $\Tsc{k}{}$ regulators, therefore, cannot be completely resolved just by addressing the details of $W(\xbj,z,\T{q}{})^{\mu \nu}$ alone, but 
instead require a treatment of $Y(\xbj,z,\T{q}{})^{\mu \nu}$. But since $Y(\xbj,z,\T{q}{})^{\mu \nu}$ is not TMD-factorizable, confronting the large-$\Tsc{k}{}$ ambiguity problem in relations like \eref{tmd2coll} leads to considerations of behavior outside of what is normally understood to be the domain of TMD physics.  

There is no 
barrier in principle to simply including a complete treatment of the $\Tsc{q}{} \sim Q$ behavior in $Y(\xbj,z,\T{q}{})^{\mu \nu}$ by using existing collinear factorization extractions for collinear PDFs and ffs. This 
has been a significant challenge in practice, however, because those large transverse momentum calculations tend to show significant tension with data~\cite{Gonzalez-Hernandez:2018ipj,Wang:2019bvb,Bacchetta:2019tcu,Moffat:2019pci} in the unpolarized case. 
It is hoped that the tension will be 
resolved by future refinements in the implementation of collinear factorization at large $\Tsc{q}{}$, and by doing this 
in parallel with TMD phenomenology. 

For applications to nucleon structure studies it might reasonably be argued that the very large perturbative transverse momentum in \eref{weighted} is not of primary interest anyway, and that the integral should be defined with a cutoff on $\T{q}{}$ at fixed and comparatively moderate momentum so as to amplify the relative contribution from truly intrinsic or nonperturbative transverse momentum.  In practice, this might be implemented by using parametrizations for TMD correlation functions, like Gaussians with fixed widths, that sharply suppress very large $\T{q}{}$ behavior. 
However, without the partonic transverse phase space growing with $Q$, it cannot be assumed automatically that evolution will follow the typical DGLAP-type behavior characteristic of most transversely integrated observables. This leads back to the question posed after \eref{evolving}. 

In the next section, I will extend the above discussion to weighted observables 
like \eref{weighted}. The large transverse momentum issue will turn out to be more interesting in this case for reasons to be explained in \sref{tmdren}. 

\section{Weighted observables}
\label{s.weighted}
The \eref{pm} approximation applied to \eref{weighted} for $n = 1$ and a component $\alpha$ of transverse momentum is analogous to \eref{reduction}. It is, 
\begin{align}
&{} \int \diff{^2 \T{q}{}} \, \Tsc{q}{}^\alpha \, W(\xbj,z,\T{q}{})^{\mu \nu}  \approx \int \diff{^2 \T{q}{}} \, \Tsc{q}{}^\alpha \, W(\xbj,z,\T{q}{})^{\mu \nu}_\text{TMD} \no
&{}=\int \diff{^2 \T{q}{}}\diff{^2 \TT{k}{1}}{} \diff{^2 \TT{k}{2}}{}  \, \Tsc{q}{}^\alpha \, \hat{W}(Q^2)^{\mu \nu}   \delta^{(2)}(\TT{k}{1} + \T{q}{} - \TT{k}{2}) 
f(x,\TT{k}{1}) D(z,z\TT{k}{2}) \no
&{}\qquad = -\hat{W}^{\mu \nu} \parz{\int \diff{^2 \TT{k}{1}}{} \, \Tsc{k}{1}^\alpha \, f(x,\TT{k}{1})  } \parz{\int \diff{^2 \TT{k}{2}}{} \, D(z,z\TT{k}{2})  } \no 
&{} \qquad \qquad + \hat{W}^{\mu \nu} \parz{\int \diff{^2 \TT{k}{1}}{}  f(x,\TT{k}{1})  } \parz{\int \diff{^2 \TT{k}{2}}{} \, \Tsc{k}{2}^\alpha \, D(z,z\TT{k}{2})  }	\,  \label{e.weights} \, .					
\end{align}
In this way, the weighted cross section gets expressed in terms of the weighted transverse moments of TMD PDFs (on the third line) and the TMD ffs (on the fourth line).
Equations analogous to \eref{pm} then connect the weighted transverse moments of TMDs to twist-3 collinear functions. For a sketch of how this works, recall that a general 
TMD PDF can be expanded in terms of polarization dependent functions, 
\begin{equation}
f(x,\T{k}{}) = f_1(x,\Tsc{k}{}) 
- \frac{\epsilon_{ij} \Tsc{k}{i} \Tsc{S}{j} }{M} f_{1T}^\perp(x,\Tsc{k}{}) + \cdots \, ,
\end{equation}
where $f_1(x,\Tsc{k}{})$ is the unpolarized 
quark TMD PDF and $f_{1T}^\perp(x,\Tsc{k}{})$ is the 
Sivers TMD PDF. The ``$\cdots$" represents other TMD functions that I do not consider here. A similar decomposition applies to the fragmentation function $D(z,z k_T)$. 
Integrating as in \eref{tmd2coll} causes contributions like 
the $f_{1T}^\perp(x,\Tsc{k}{})$ term to 
vanish due to the odd factor of $\Tsc{k}{i}$.
However, in the integral in parentheses on the third line of \eref{weights} it does not vanish. 
Instead it produces a factor proportional to the integral 
\begin{equation} \int \diff{^2\T{k}{}} \Tscsq{k}{} f_{1T}^\perp(x,\Tsc{k}{}) \, . 
\label{e.weightedint}
\end{equation}
The asymptotic large $\Tsc{k}{}$ behavior of $f_{1T}^\perp(x,\Tsc{k}{})$ is $1/k_T^4$, so here again there is an ultraviolet $k_T$ problem.
Completely removing the large $k_T$ ambiguity means including both $W(\xbj,z,\T{q}{})^{\mu \nu}_\text{TMD}$ and $Y(\xbj,z,\T{q}{})^{\mu \nu}$ in the weighted integral, 
analogously to \eref{fullintegral}. 

However, if one momentarily sets aside the treatment of ultraviolet divergences in the operator definitions of correlation functions, it is possible 
to show that the extra power of transverse momentum translates into a derivative and then to derive an integral relation analogous to \eref{tmd2coll} but with a twist-3 collinear function on the right-hand side,
\begin{equation}
\int \diff{^2\T{k}}{} \frac{\Tscsq{k}{}}{M^2} f_{1T}^{\perp}(x,\Tsc{k}{}) =  - \frac{1}{M} T(x) 
\, .
\label{e.basic0}
\end{equation}
The $T(x)$ on the right side is a twist-3 quark-gluon-quark collinear correlation function often called the Efremov-Teryaev-Qiu-Sterman (ETQS) function~\cite{Efremov:1984ip,Qiu:1991pp,Qiu:1991wg,Qiu:1998ia}. Equation~\eqref{e.basic0} was first derived in Ref. \citen{Boer:2003cm}.
The minus sign on the right-hand side of \eref{basic0}
is consistent with a definition for the TMD PDF with a future pointing Wilson line, as is needed for SIDIS. 

In the next section, I will contrast 
the different types of violations that can arise from large $\Tsc{k}{}$ in \eref{tmd2coll} and \eref{basic0}.

\section{Transverse momentum regulators and renormalization}
\label{s.tmdren}
  
Ultraviolet divergences in the transverse momentum integrals of the previous sections create the possibility for violations of relations 
like \eref{tmd2coll} and \eref{basic0}. 
Addressing this requires precise statements of the operator definitions used for correlation functions, including whether 
operators are renormalized or bare and whether operator products are defined with renormalization or with cutoffs. Note that renormalization and cutoff regularization are not exactly the same.\cite{Collins:2003fm}. In PDF renormalization (see, e.g., Sec.~8.3 of Ref.~\citen{Collins:2011zzd}), a bare PDF $f_0$ is defined first, with bare fields and parameters. Then, a renormalized PDF is defined by applying a generalized renormalization factor $Z$,
\begin{equation}
f^\text{renorm} \equiv Z \otimes f_0 \, , 
\end{equation}
where $\otimes$ is the usual integral convolution in longitudinal momentum fraction. In a scheme like $\msbar$, $Z$ implements the subtraction of ultraviolet poles. 
A renormalized PDF defined in this way is not generally reproduced by integrating a TMD PDF up to a cutoff. (The TMD PDFs are defined with their own, separate renormalization procedures.) 

In the discussions that follow, TMD and collinear PDFs and any of the other correlation functions should be understood to be defined in any of the usual ways relevant to applications, with renormalized operators and renormalized operator products for collinear correlation functions. The treatment of lightcone divergences and Wilson lines in TMD functions should be understood to follow any of the now standard approaches~\cite{Collins:1981uw,Collins:2011zzd,Diehl:2015uka,Aybat:2011ge,Rogers:2015sqa,Stewart:2009yx,Becher:2010tm, Becher:2011xn, Becher:2012yn, GarciaEchevarria:2011rb, Echevarria:2012js, Echevarria:2014rua, Chiu:2012ir,Li:2016axz}, although the precise details of this particular issue will be unimportant for the discussion below. For collinear PDFs and other collinear correlation functions, there are a number of advantages to using renormalized operator matrix elements as definitions, with standard renormalization prescriptions like $\overline{\text{MS}}$. Firstly, they possess desirable features like the automatic cancellation of lightcone divergences~\cite{Collins:2003fm}. 
Secondly, properties like equations of motion and sum rules for composite operators are exactly valid only in a limited number of such schemes~\cite{Collins:1984xc}, including \msbar. Finally, the use of schemes like $\overline{\text{MS}}$ is already pervasive in existing phenomenological treatments of collinear functions that include higher order QCD. Thus, all collinear functions will be assumed to be treated in this way below.

Using integrals like \eref{tmd2coll} and the left side of \eref{basic0} with actual TMD correlation functions that have been extracted from phenomenological fitting requires cutting off or suppressing the large $\T{k}{}$ region while leaving the small and physically more relevant $\T{k}{}$ contribution unchanged. This can be done smoothly, for example by using a Gaussian parametrization for the large $\T{k}{}$ tail, or with a sharp cutoff, though the general observations below are independent of such choices so I will use hard cutoffs for simplicity. 

In \eref{tmd2coll}, for example, the 
integral on the left side is to be defined 
with a large transverse momentum cutoff $k_c$. If the right side is the standard renormalized unpolarized quark PDF, then the size of a violation of \eref{tmd2coll} is measured by the quantity
\begin{equation}
\Delta f
\equiv \parz{\pi \int_0^{k_c^2} \diff{\Tscsq{k}{}}{} \,   
f_1(x,\Tsc{k}{})} - f(x;\mu) \, . 
\label{e.deltdef1}
\end{equation}
It is straightforward to verify that calculations of 
$\Delta f$ follow a typical collinear factorization pattern. 
Namely,
\begin{equation}
\Delta f = 
\mathcal{C} \parz{x,\alpha_s(\mu)} \otimes f(x;\mu) + \order{\frac{\Lambda^2_{\rm QCD}}{k_c^2}} \,  ,\label{e.yuk}
\end{equation}
where $\mathcal{C} \parz{x,\alpha_s(\mu)}$ is a hard coefficient that starts at order 
$\alpha_s(\mu)$ if $k_c \sim \mu$. Most relevantly here, 
\begin{equation}
\Delta f = \order{\alpha_s(k_c)} \, , 
\label{e.deltdef2p}
\end{equation}
so that asymptotic freedom ensures $\Delta f \to 0$ when both
$k_c$ and $\mu$ are fixed to some hard scale $Q$ and $Q/\Lambda_\text{QCD} \to \infty$. 
This is consistent with the natural intuition that, because the 
violation of \eref{tmd2coll} is from a hard $\Tsc{k}{}$ tail, the effect of a nonzero $\Delta f$ is likely to be perturbatively suppressed.

Given these observations regarding the unpolarized correlation functions, it is natural to expect 
something similar in relations involving polarization dependent functions such as \eref{basic0}. 
The analog of \eref{deltdef1}, for example, is
\begin{equation}
\Delta f_{1T}^\perp 
\equiv \parz{\pi \int_0^{k_c^2} \diff{\Tscsq{k}{}}{} \frac{\Tscsq{k}{}}{M^2} f_{1T}^{\perp}(x,\Tsc{k}{})} 
 + \frac{1}{M} T(x) \, . \label{e.deltdef2}
\end{equation}
The conclusion of Ref.~\citen{Qiu:2020oqr}, however, is 
that the analogous arguments lead not 
to something like \eref{deltdef2p}, but rather to 
 \begin{equation}
 \Delta f_{1T}^\perp \sim \alpha_s(k_c)^2 \ln^2 \parz{\frac{k_c^2}{m^2}} \, , \label{e.deltTperp}
 \end{equation}
up to overall factors, where $m$ is a small nonperturbative mass scale, roughly the size of $\Lambda_\text{QCD}$ (see Eq.~(13) of Ref.~\citen{Qiu:2020oqr} and the surrounding discussion there). 

The asymptotic behavior of the strong coupling is 
$\alpha_s(k_c) \sim 1/\ln\parz{k_c/m}$, so the asymptotic freedom of QCD does not lead to a suppression of \eref{deltTperp}. It is less obvious in the polarization case, therefore, that neglecting  violations of \eref{basic0} is a good approximation.

\section{Summary}
\label{s.summary}

Equation~\eqref{e.deltTperp} implies a stronger 
ambiguity for the definition of the integral on the 
left-hand side of \eref{basic0} than might be expected. Since it is not 
a normal type of perturbative correction, it can have a potentially greater impact on observables than the analogous quantity in the unpolarized case, \erefs{yuk}{deltdef2p}.

Reference~\citen{Qiu:2020oqr} argued that the resolution to this problem can be guided by the type of physics of greatest interest for a particular application. Thus, for instance, applications to nucleon structure might use a fixed and relatively low regulator on transverse momentum in \eref{weightedint}. In such cases, Ref.~\citen{Qiu:2020oqr} proposes taking the relation in \eref{deltTperp} to \emph{define} a scheme for the ultraviolet behavior of $T(x)$. 
The advantage of such a scheme is that 
it preserves the parton model picture embodied by relations like \eref{tmd2coll} and \eref{basic0} along with its applications, some of which were mentioned in the introduction. Furthermore, it avoids having to directly address the question of the $Y$-term correction by eliminating the $\Tsc{q}{} \sim Q$ contribution. 

However, this low transverse momentum cutoff, along with \eref{deltTperp}, means this definition likely does not preserve the normal renormalization group evolution of $T(x)$. (Note that~\eref{deltTperp} implies that different numbers of large logarithms are included in the cut off TMD PDF and the renormalized collinear function.)  Instead, TMD evolution should be implemented first on the TMD PDF inside the integrand of \eref{deltdef2}. This will be relevant to future refinements to the treatment of evolution in applications to phenomenology such as Ref.~\citen{Cammarota:2020qcw}. More work along these lines is needed.

\section*{Acknowledgments}

I thank F.~Aslan, L.~Gamberg, J.-W.~Qiu, N.~Sato, and B.~Wang for many useful discussions related to the above topics. 
This work was supported by the U.S. Department of Energy, Office of Science, Office of Nuclear Physics, under Award Number DE-SC0018106, and by the U.S. Department of Energy contract DE-AC05-06OR23177, under which Jefferson Science Associates, LLC, manages and operates Jefferson Lab.

\appendix

\bibliographystyle{ws-ijmpa}
\bibliography{bibliography}
\end{document}